# Daily Deals: Prediction, Social Diffusion, and Reputational Ramifications

John W. Byers[*]
Computer Science Dept.
Boston University

Michael Mitzenmacher[†]
School of Eng. Appl. Sci.
Harvard University

Georgios Zervas[*]
Computer Science Dept.
Boston University

## ABSTRACT

Daily deal sites have become the latest Internet sensation, providing discounted offers to customers for restaurants, ticketed events, services, and other items. We begin by undertaking a study of the economics of daily deals on the web, based on a dataset we compiled by monitoring Groupon and LivingSocial sales in 20 large cities over several months. We use this dataset to characterize deal purchases; glean insights about operational strategies of these firms; and evaluate customers' sensitivity to factors such as price, deal scheduling, and limited inventory. We then marry our daily deals dataset with additional datasets we compiled from Facebook and Yelp users to study the interplay between social networks and daily deal sites. First, by studying user activity on Facebook while a deal is running, we provide evidence that daily deal sites benefit from significant word-of-mouth effects during sales events, consistent with results predicted by cascade models. Second, we consider the effects of daily deals on the longer-term reputation of merchants, based on their Yelp reviews before and after they run a daily deal. Our analysis shows that while the number of reviews increases significantly due to daily deals, average rating scores from reviewers who mention daily deals are 10% lower than scores of their peers on average.

## 1. INTRODUCTION

Groupon and LivingSocial are websites offering various deals-of-the-day, with localized deals for major geographic markets. Groupon in particular has been one of the fastest growing Internet sales businesses in history, with tens of millions of registered users and 2011 sales expected to exceed 1 billion dollars.

We briefly describe how daily deal sites work; additional details relevant to our measurement methodology will be given subsequently. In each geographic market, or city, there are one or more deals of the day. Generally, one deal in each market is the featured deal of the day, and receives the prominent position on the primary webpage targeting that market. The deal provides a coupon for some product or service at a substantial discount (generally 40-60%) to the list price. Deals may be available for one or more days. We use the term *size* of a deal to represent the number of coupons sold, and the term *revenue* of a deal to represent the number of coupons multiplied by the price per coupon. Groupon retains approximately half the revenue from the discounted coupons [10], and provides the rest to the seller, as does LivingSocial. Deals each have a minimum threshold size that must be reached for the deal to take hold, and sellers may also set a maximum threshold size to limit the number of coupons sold.

Daily deal sites represent a change from recent Internet advertising trends. While large-scale e-mail distributions for sale offers are commonplace (generally in the form of spam) and coupon sites have long existed on the Internet, Groupon and LivingSocial have achieved notable success with their emphasis on higher quality localized deals, as well as their marketing savvy both with respect to buyers and sellers (merchants). This paper represents an attempt to gain insight into the success of this business model, using a combination of data analysis and modeling.

The contributions of the paper are as follows:

- We compile and analyze datasets we gathered monitoring Groupon over a period of six months and LivingSocial over a period of three months in 20 large US markets. Our datasets will be made publicly available (on publication of this paper).

- We consider how the price elasticity of demand, as well as what we call "soft incentives", affect the size and revenue of Groupon and LivingSocial deals empirically. Soft incentives include deal aspects other than price, such as whether a deal is featured and what days of the week it is available.

- We study the predictability of the size of Groupon deals, based on deal parameters and on temporal progress. We show that deal sizes can be predicted with moderate accuracy based on a small number of parameters, and with substantially better accuracy shortly after a deal goes live.

- We examine dependencies between the spread of Groupon deals and social networks by cross-referencing our Groupon dataset with Facebook data tracking the frequency with which users "like" Groupon deals. We offer evidence that propagation of Groupon deals is consistent with predictions of social spreading made by cascade models.

- We examine the change in reputation of merchants based on their Yelp reviews before and after they run a Groupon deal. We find that reviewers mentioning daily deals are significantly more negative than their peers on average, and the volume of their reviews materially lowers Yelp scores in the months after a daily deal offering.

We note that we presented preliminary findings based on a single month of Groupon data that focused predominantly on the issue of soft incentives in a technical report [3]. The current paper enriches that study in several ways, both in its consideration of LivingSocial as a comparison point, and especially in our use of social network data sources, such as Facebook and Yelp, to study deal sites. Indeed, we believe this use of multiple disparate data sources, while not novel as a research methodology, appears original in this context of gaining insight into deal sites.

[*]E-mail: {byers, zg}@cs.bu.edu. Supported in part by Adverplex, Inc. and by NSF grant CNS-1040800.

[†]E-mail: michaelm@eecs.harvard.edu. Supported in part by NSF grants CCF-0915922 and CNS-0721491, and in part by a research grant from Yahoo!

Before continuing, we acknowledge that a reasonable question is why we gathered data ourselves, instead of asking Groupon for data; such data (if provided) would likely be more accurate and possibly more comprehensive. We offer several justifications. First, by gathering our own data, we can make it public, for others to use and to verify our results. Second, by relying on a deals site as a source for data, we would be limited to data they were willing to provide, as opposed to data we thought we needed (and was publicly available). Gathering our own data also motivated us to gather and compare data from multiple sources. Finally, due to fortuitous timing, Groupon's recent S-1 filing [10] allowed us to validate several aggregate measures of the datasets we collected.

**Related Work on Daily Deals:** To date, there has been little previous work examining Groupon and LivingSocial specifically. Edelman et al. consider the benefits and drawbacks of using Groupon from the side of the merchant, modeling whether the advertising and price discrimination effects can make such discounts profitable [9]. Dholakia polls businesses to determine their experience of providing a deal with Groupon [8], and Arabshahi examines their business model [2]. Several works have studied other online group buying schemes that arose before Groupon, and that utilize substantially different dynamic pricing schemes [1, 12]. Ye et al. recently provide a stochastic "tipping point" model for sales from daily deal sites that incorporates social network effects [21]. They provide supporting evidence for their model using a Groupon data set they collected that is similar to, but less comprehensive, than ours, but they do not measure social network activity.

## 2. THE DAILY DEALS LANDSCAPE

In this section, we describe the current landscape of daily deal sites exemplified by Groupon and LivingSocial. We start by describing the measurement methodology we employed to collect longitudinal data from these sites, and provide additional background on how these sites operate. We then describe basic insights that can be gleaned directly from our datasets, including revenue and sales broken out by week, by deal, by geographic location, and by deal type. Moving on, we observe that given an offering, daily deal sites can optimize the performance of the offering around various parameters, most obviously price, but also day-of-week, duration, etc. We explore these through the lens of our datasets.

### 2.1 Measurement Methodology

We collected longitudinal data from the top two group deal sites, Groupon and LivingSocial, as well as from Facebook and Yelp. Our datasets are complex and we describe them in detail below.

#### 2.1.1 Deal data

We collected data from Groupon between January 3rd and July 3rd, 2011. We monitored – to the best of our knowledge – all deals offered in 20 different cities during this period. Our criteria for city selection were population and geographic distribution. Specifically, our list of cities includes: Atlanta, Boston, Chicago, Dallas, Detroit, Houston, Las Vegas, Los Angeles, Miami, New Orleans, New York, Orlando, Philadelphia, San Diego, San Francisco, San Jose, Seattle, Tallahassee, Vancouver, and Washington DC. In total, our data set contains statistics for 16,692 deals.

Each Groupon deal is associated with a set of features: the deal description, the retail and discounted prices, the start and end dates, the threshold number of sales required for the deal to be activated, the number of coupons sold, whether the deal was available in limited quantities, and if it sold out. Each deal is also associated with a category such as "Restaurants", "Nightlife", or "Automotive". From these basic features we compute further quantities of interest such as the revenue derived by each deal, the deal duration, and the percentage discount.

With each Groupon deal, we collected intraday time-series data which monitors two time-varying parameters: cumulative sales, and whether or not a given deal is currently *featured*. To compile these time-series, we monitored each deal in roughly ten-minute intervals and downloaded the value of the sales counter. Occasionally some of our requests failed and therefore some gaps are present in our time-series data, but this does not materially affect our conclusions. The second parameter we monitored was whether a deal was featured or not, with featured deals being those deals that are presented in the subject line of daily subscriber e-mails while being given prominent presentation in the associated city's webpage. For example, visiting `groupon.com/boston`, one notices that a single deal occupies a significant proportion of the screen real-estate, while the rest of the deals which are concurrently active are summarized in a smaller sidebar.

Although Groupon has a public API[1] through which one can obtain some basic deal information, we decided also to monitor the Groupon website directly. Our primary rationale was that certain deal features, such as whether a link to reviews for the merchant offering the deal was present, were not available through the Groupon API. We used the API to obtain a category for each deal and to validate the sales data we collected. Observed discrepancies were infrequent and small: we used the API-collected data as the ground truth in these cases. We did not use the API to collect time-series data.

We collected data from LivingSocial between March 21st and July 3rd, 2011 for the same set of 20 cities. In total, our LivingSocial dataset contains 2,609 deals. LivingSocial deals differ from their Groupon counterparts in that they have no tipping point, and in that they do not explicitly indicate whether they are available in limited quantities (although they do sell out occasionally). LivingSocial runs two types of deals: one featured deal per day, and a secondary "Family Edition" deal, which offers family-friendly activities, and receives less prominent placement on the LivingSocial website. For LivingSocial deals we only collected data on their outcomes; we did not collect time-series data.

#### 2.1.2 Facebook data

Both Groupon and LivingSocial display a Facebook Like button for each deal, where the Like button is associated with a counter representing the number of Facebook users who have clicked the button to express their positive feed-

---

[1] http://www.groupon.com/pages/api

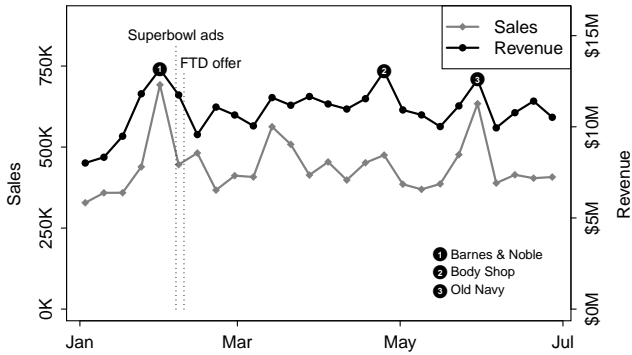
(a) Groupon

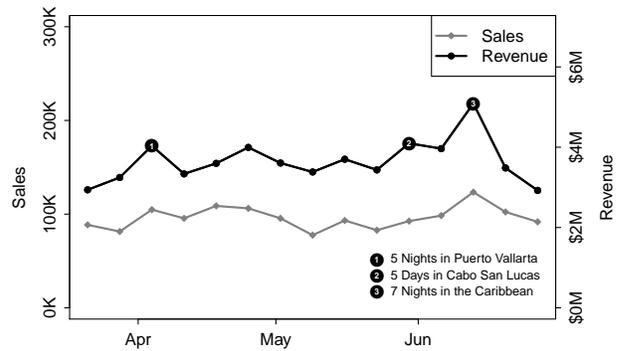
(b) LivingSocial

Figure 1: Weekly revenue and sales in 20 selected cities (2011).

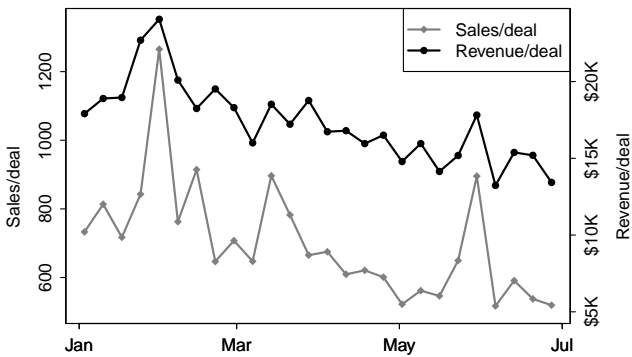
(a) Groupon

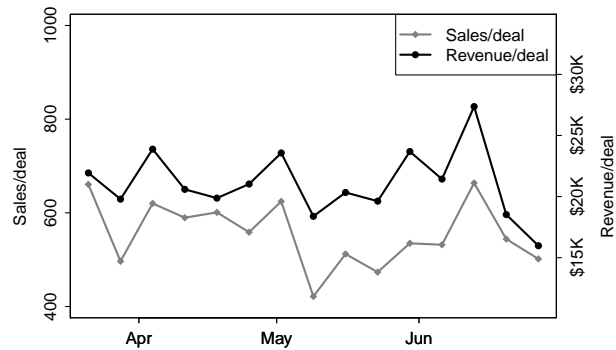
(b) LivingSocial

Figure 2: Revenue and coupons sold per deal week-over-week.

back. We refer to the value of the counter as the number of *likes* a deal has received, and we collected this value for each Groupon and LivingSocial deal in our dataset.

As a technical aside, we mention that Groupon and LivingSocial have different implementations of the Facebook Like button that necessitated our collecting data from them in different ways. Within each deal page, Groupon embeds code that dynamically renders the Facebook Like button. It does so by sending Facebook a request that contains a unique identifier associated with the corresponding deal page. We extracted the unique identifier from Groupon deal pages and directly contacted Facebook to obtain the number of likes for every deal. LivingSocial instead hard-codes the Like button and its associated counter within each page. As we could not obtain the identifier associated with each LivingSocial deal, we could not query Facebook to independently obtain the number of likes, and thus we collected the hard-coded number from LivingSocial deal pages.

### 2.1.3 Yelp data

Groupon occasionally displays reviews for the merchant offering the deal in the form of a star-rating, as well as selected reviewer comments. The reviews are sourced from major review sites such as Yelp, Citysearch, and TripAdvisor. For Groupon deals that were linked to Yelp reviews, we collected the individual reviewer ratings and comments left by customers on Yelp. We collected this dataset during the first week of September 2011. In total, our dataset contains 56,048 reviews, for 2,332 merchants who ran 2,496 deals on Groupon during our monitoring period. Yelp has implemented one measure to discourage the automated collection of reviews which directly affected our data collection: it hides a set of user reviews for each merchant. To see the hidden reviews, one has to solve a CAPTCHA. We did not attempt to circumvent CAPTCHAs, and we do not know whether the hidden reviews are randomly selected, or are selected by other criteria. Since Yelp reports the total number of reviews available per merchant, we ascertained that approximately 23% of all reviews for merchants in our dataset were hidden from our collection.

## 2.2 Operational Insights

Figure 1 serves as an overview of insights we are able to gain using our dataset. It displays the weekly revenue as well as the weekly sales of coupons across all 20 cities we monitored for Groupon and LivingSocial, respectively. Notably, while both Groupon and LivingSocial are widely regarded as companies enjoying extremely rapid growth, our first takeaway from these plots is that sales and revenue in these 20 established markets are relatively flat across the time period. We conjecture that much of the reported growth is in newer markets.

By happenstance, Groupon's recent S-1 filing [10] provided financial information that allowed us to validate some of the aggregate revenue data that we collected indepen-dently. For example, the filing states that in the three months ending March 31, 2011, Groupon had sold 950,689 deals in Chicago earning $21.5M in revenues. Our dataset accounts for 967,244 deals sold and $21.3M in revenues. For the same period in Boston, the filing reports 388,178 deals sold for total revenues of $9.3M compared to 362,823 deals and $8.7M in our dataset. In both cases, our observations closely match what the company reported. The small differences may have arisen for several reasons: we did not monitor all revenue generating activities (such as direct backchannels provided to merchants); some deals were offered at multiple prices but we only monitored the main one; our accounting methods might differ from Groupon's; our inability to account for refunds; and our monitoring infrastructure may have overlooked some deals.

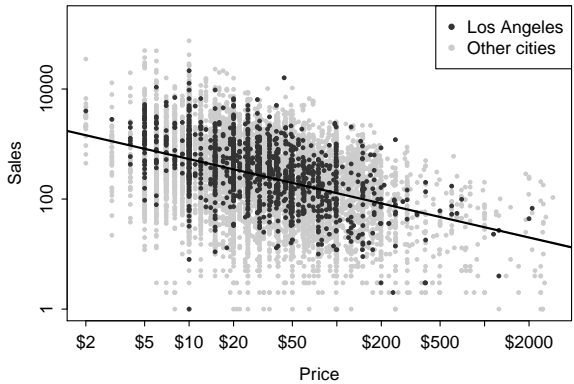 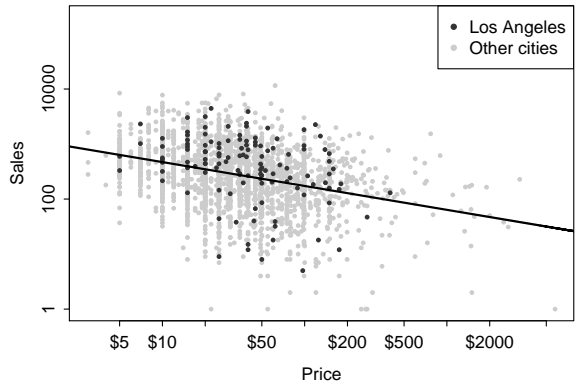

(a) Groupon, slope = −0.62   (b) LivingSocial, slope = −0.43

Figure 3: Deal size by price (in logarithmic scale.) Black dots highlight deals in LA; grey dots are used for other cities. A trend line fitted using OLS regression for all data is also shown.

Outliers in our plots are not due to especially strong performance in local markets, but instead seem to be generated by large national deals. The three most significant outliers for both Groupon and LivingSocial each correspond to a large national offer. Representative deals noted in Figure 1 include $10 for $20 in Barnes & Noble merchandise during the week of 1/31 for Groupon and 7 nights in the Caribbean during the week of 6/13 for LivingSocial.

The Groupon plot is annotated with two significant events from the February time period that mark an unusual decline in revenues. On February 6th, Groupon's Super Bowl commercials aired, and were found offensive by many people; Groupon apologized on their official blog and pulled the controversial ad campaign [11]. Subsequently, from February 9th to February 11th, in advance of Valentine's Day, Groupon offered a nationwide "$20 for $40 Worth of Flowers" coupon in partnership with FTD Flowers. Shortly thereafter, customers started realizing that when visiting the FTD Flowers website via Groupon they were shown much higher prices for the same products compared to customers who visited the FTD site directly. Furthermore, coupon purchasers were only allowed to use their coupons on the more expensive version of the product. This deal was widely perceived as a "bait and switch" scheme; Groupon offered disgruntled customers refunds[2] [18]. The recovery that we observe in the revenue indicates that this shock may only have had a short-term effect on Groupon's business. Similar short-term effects have been observed in the time-series of YouTube video views after exogenous shocks [7].

A different phenomenon is captured in Figure 2, where we plot week-over-week sales and revenue on a *per-deal* basis. For Groupon, sales and revenue per deal in the 20 cities we monitored peaked in February 2011 and have trended steadily downward since. For LivingSocial, we observe much greater variability, obscuring any underlying trend, but per-deal sales and revenue appear flat or declining. While these trends could potentially point to underlying fragility in the daily deals business model (possibly the best revenue-producing merchants in a geographical area are becoming exhausted), more benign explanations exist. The trend could reflect a change in operational strategy to broaden the base of featured deals available to subscribers at any given time, for example, to provide better personalization of deals to subscribers. However, revenue growth in established markets appears to be a potential challenge facing daily deal sites.

## 2.3 Pricing and Other Incentives

We now consider a range of factors that influence the number of coupons sold and that daily deal sites can control. Our primary focus is on Groupon. As daily deal sites have built their business around deeply discounted offers, one might expect that the discounted price associated with a deal would be the primary purchase incentive. Indeed, in the log-log scatterplots of deal sizes vs. deal prices depicted in Figure 3, the trend lines fitted using ordinary least-squares (OLS) regression indicate that the logarithm of price and the logarithm of sales are roughly linearly related. While there is a large amount of variance within individual price points, by controlling for other features, such as restricting attention to deals in Los Angeles (black points), the trend becomes clearer.

A closely related deal feature is the magnitude of the discount, which Groupon displays prominently in advertising a deal. While we conjecture that customers are sensitive to this quantity, most discounts presented fall in a relatively narrow range (three-quarters of all Groupon deals are discounted by 40 to 60%). This has the effect of making the list price highly correlated (0.90 correlation coefficient) with

---

[2]Since we cannot account for these refunds, the reported sales and revenues of the week starting on February 7 are an overestimate.

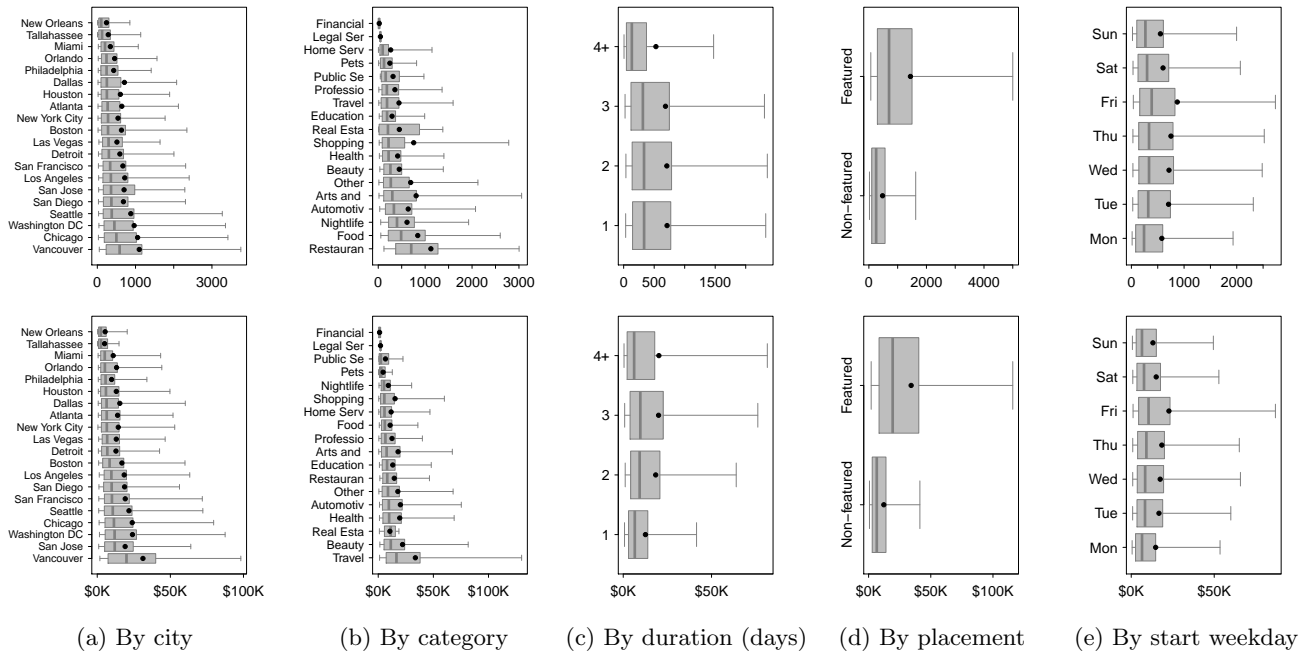

(a) By city  (b) By category  (c) By duration (days)  (d) By placement  (e) By start weekday

Figure 4: Groupon deal sizes (top row) and deal revenues (bottom row) are shown on the $x$−axis. On the $y$−axis deals are broken down by various deal features. The outer whiskers mark the 5% and 95% quantiles of the deal size, the sides of each grey rectangle the 1st and 3rd quartiles, the solid bar the median, and the solid dot the mean.

| Duration (days) | 1 | 2 | 3 | 4+ |
|---|---:|---:|---:|---:|
| Mean price | $28 | $42 | $56 | $139 |
| Mean sales | 712 | 707 | 685 | 529 |
| Mean revenue | $12,576 | $18,375 | $20,010 | $20,189 |
| Number of deals | 5,464 | 5,745 | 3,877 | 1,606 |

Table 1: Groupon deals by duration.

|  | Featured | Non-featured |
|---|---:|---:|
| Mean sales | 1,443 | 475 |
| Mean revenue | $34,181 | $12,241 |
| Number of deals | 3,644 | 13,048 |

Table 2: Groupon deals by placement.

the discount price and therefore the discount is a weak distinguishing feature once price has been taken into account.

Deal sites can also control the length of time a deal runs. As shown in Table 1, while deal size varies minimally with duration, revenue increases, suggesting perhaps that Groupon is attempting to hit *sales*, instead of *revenue*, targets. Expensive deals may generally sell less quickly, and must be allowed more time to achieve the same sales goal.

At any point in time, and for each geographic market, one deal among the set of all available deals is featured by Groupon. Featured deals receive prominent placement both on the Groupon website and in the daily email that customers receive. In our dataset, 22% of deals were featured. The impact of featured placement is significant: the mean sales and revenue for featured deals are well in excess of twice the corresponding quantities for other deals. The effects of featuring a deal are summarized in the Table 2. However, we cannot assume that these outcomes are entirely causal effects from featuring a deal, as Groupon naturally has an incentive to feature those deals that will drive the most sales. We investigated whether certain *categories* of deals as a whole were featured more frequently. Table 3 breaks down categories with at least 100 deals by the percentage of deals within each that was featured.

Another distinguishing characteristic is inventory size: some deals are available only in limited quantities. Groupon lets its customers know which deals are limited, but does not display the number of available units (while some competitors of Groupon display this prominently). Approximately 31% of all deals in our dataset are available in limited numbers, with wide variation across categories. Approximately 50% of all "Travel" deals and 39% of all "Arts and Entertainment" deals were available in limited quantities, as one might expect for these types of deals. Surprisingly, only about 18% of the limited deals in our dataset sold out. It is not that these deals are intrinsically less attractive: limited deals outperformed unlimited deals on average by 11% more coupon sales and 27% more revenue. It is possible that merchants (or Groupon) artificially limit deals as a strategy to exert pressure to customers, making them more likely to purchase on the spur of the moment.

Groupon also has a choice as to the days of the week that it schedules each offer. Figure 4e breaks down deals by the day of the week on which they began. Even though the differences are not striking, it appears that deals starting in the beginning of the week produce less revenue, and deals starting on Friday produce the most. One possible explanation is that on Fridays, Groupon starts more multi-day deals that span the weekend. For example, 45% of three-day deals start on a Friday. However, alternative explanations, such as that the best deals are run on Friday or that people are more likely to buy on Friday, may also apply.

|  | Featured (%) | # of deals |
| --- | --- | --- |
| Arts & Entert. | 31% | 2,932 |
| Automotive | 16% | 373 |
| Beauty & Spas | 17% | 2,476 |
| Education | 24% | 508 |
| Food & Drink | 17% | 1,193 |
| Health & Fitness | 18% | 2,092 |
| Home Services | 13% | 370 |
| Nightlife | 25% | 288 |
| Other | 24% | 327 |
| Prof. Services | 18% | 673 |
| Restaurants | 24% | 2,794 |
| Shopping | 18% | 2,176 |
| Travel | 35% | 404 |

Table 3: A breakdown of featured deals by category.

Deals are said to be "on" when they surpass a sales threshold defined by Groupon, possibly in conjunction with the merchant. That is, for customers to get a discount, a minimum number of them must commit to the deal. In theory, this could drive a group dynamic whereby customers encourage their friends to buy a coupon to reach the threshold. However, current deal thresholds are very low. For example, in our dataset, the mean threshold to total sales ratio is approximately 19%. Deals that surpassed their threshold did so on average just after 8am, and only 2% of deals with thresholds failed to reach them. Like the relationship between price to deal size, the relationship between threshold and deal size also appears to be linear when plotted in logarithmic scale.

Deals can also be classified by two features that Groupon does not have direct control over: their geographic market, and their category. Figure 4a shows that while overall deal sizes vary considerably across cities, and not always in proportion to city population. This is in part due to localization: Long Island has a separate Groupon deal stream from New York; similarly Los Angeles is split into multiple sub-areas. As for deal categories, Figure 4b suggests that deals that are the most lucrative in terms of revenue for Groupon are not the most popular with their customers. For example, while "Travel" deals produce fewer sales than most other categories on average, they produce the most revenue; conversely "Restaurants" deals are the most popular, but produce much less revenue.

Finally, it is not especially surprising, but still notable that we observe heavy-tailed behavior throughout Figure 4, with the mean deal producing many more sales and much more revenue than the median in essentially all cases.

## 3. MODELING DEAL OUTCOMES

Having considered various deal features and their individual correlation with deal size and revenue individually, we now consider these characteristics collectively using regression. Our goal is both to better quantify the dependence of deal outcomes on the various features, and to determine if such models are sufficiently accurate to predict the outcomes of future deals. The model we use for Groupon deals, noting that all logarithms are base $e$, is:

$$\begin{aligned} \log q &= \beta_0 + \beta_1 \log p + \beta_2 \log t + \beta_3 d + \beta_4 f + \beta_5 l \\ &+ \bar{\beta}_6 \mathbf{w} + \bar{\beta}_7 \mathbf{c} + \bar{\beta}_8 \mathbf{g} \end{aligned} \quad (1)$$

where $q$ stands for the deal size, $p$ for the coupon price, $t$ for the threshold, $d$ for whether the deal is run for multiple days or not, $f$ for whether the deal is featured or not, and $l$ for whether the deal inventory is limited or not. The values of $p$ and $t$ are centered to their corresponding medians (25 in both cases). This allows for a more intuitive interpretation of the regression's intercept but does not otherwise affect our results. The parameters $\mathbf{w}$, $\mathbf{c}$ and $\mathbf{g}$ are dummy-coded vectors representing the starting day of the week, category, and city relative to notional reference levels; their corresponding coefficients are also vectors. Dummy-coding refers to using binary vectors to encode categorical variables, where a variable that can take on $k$ distinct values is encoded using a binary vector of length $k-1$ where at most one entry is set to one. We also fitted a similar log-log model to LivingSocial deals with similar results to those we report below.

The exact form the model takes upon is motivated by the observations of Section 2.3: $\log p$ and $\log t$ are well modeled as having a linear relationship to $\log q$, while the rest of the variables in our model are either boolean or categorical in nature. Given the high correlation of the list price to the discounted price, we have chosen to exclude it from the model to avoid introducing multicollinearity. Also, since most multi-day deals last for two days, and there is little variance in the number of sales among multi-day deals, we have chosen (after experimentation) to encode duration as a boolean feature.

We fitted the model using ordinary least squares (OLS) regression. The parameter estimates, their standard errors, and their significance levels are shown in Table 7a in the appendix. A similar model of LivingSocial deals appears in Table 7b. We focus the presentation of our results on Groupon.

The intercept of the model is the unconditional expected mean of the logarithm of the size of a deal. For example, the expected mean of the log-size of a deal in the "Other" category, priced at $25, with a threshold of 25, not featured, with unlimited inventory, starting on Monday, and running for a single day in Atlanta, is 5.19. Equivalently, the expected *geometric* mean of the size of the same deal is $e^{5.19} \approx 179$.

The coefficient of $\log p$ is particularly interesting because its value is the *point-price elasticity of demand* for coupons. To see this, recall that the point-elasticity $\eta_p$ is defined as:

$$\eta_p = (\partial q/q)/(\partial p/p) = \partial \log q / \partial \log p = \beta_1. \quad (2)$$

For deal size, the point-price elasticity given by our regression model is $-0.48$. Intuitively, this means that for a 1% increase in price, we expect a (roughly[3]) 0.48% decrease in demand. Since $\eta_p > -1$ the demand is said to be *inelastic*. This matches our intuition: coupons already represent heavy discounts and as such, changes in price should have a relatively less significant impact upon demand.

The coefficients of non-log-transformed variables represent differences in the expected means of log-sales, and their ex-ponentiated values represent multiplicative increases (or decreases) in the expected geometric mean of sales. For example, the expected ratio of the geometric means of sales for multi-day deals to single-day deals is $e^{0.22} \approx 1.25$. A simple interpretation is that by running a deal for more than a day, we expect a 25% increase in sales. The effect of featuring a deal is, as anticipated, far greater: featured deals are expected to perform 141% better than their non-featured counterparts. Limiting a deal, in the presence of all the other parameters in our model, does not have a statistically significant effect.

---
[3] The point-price elasticity relates infinitesimal percentage changes in price to percentage changes in demand. For any small enough percentage change in price the estimated percentage change in sales will be reasonably accurate.

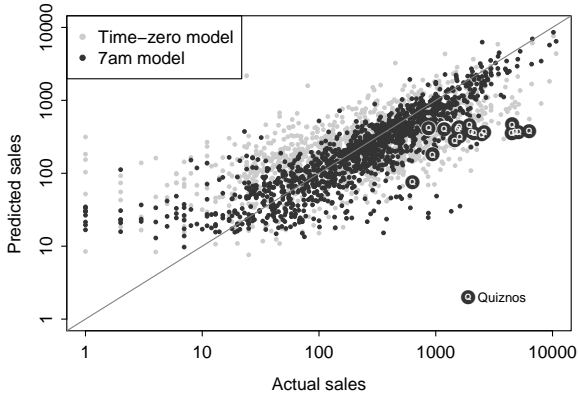

Figure 5: Actual vs. predicted sales for a test set of Groupon deals, in log-log scale for our plain regression model, as well as the the model incorporating early morning sales.

Overall, the $F$-statistic (345) and the $p$-value ($< 2 \cdot 10^{-16}$) of the model indicate that we can reject the null hypothesis (that there is no relation between sales and any of the explanatory variables in our model) with high confidence. However, the $R^2$ (0.49) value of the model suggests moderate predictive power. We tested this against a test set of 1,522 deals that ran in the same 20 cities between August 18th and 31st. Figure 5 shows the number of sales our model predicts against their actual sales. As anticipated by the $R^2$ value of our model, our predictions are generally well correlated with actual sales, with noticeably large errors for certain individual deals.

Not surprisingly, our model consistently overestimates weak deals with a very small number of actual sales, and tends to underestimate the strongest deals, indicating that there are additional factors at work not captured by our model. One such factor we study in Section 4.1 is how social networks can influence deal sales. Another possible influence on purchasing decisions could be merchant reputations, which we consider below, and in the other direction (how Groupon offers affect reputation) in Section 4.2. One clear major influence is that some deals are national in scope, and our largest underestimates appear to be for these deals. For example, as shown in in Figure 5, some of the larger errors in underestimating deal size correspond to a national deal for Quiznos sandwiches. Incorporating these additional refinements to improve the predictive power of our models is future work.

We consider two variations on our basic model:

**Early-morning sales:** We have incorporated first-day sales at 7am from our time-series dataset into our basic model. Our reasoning is that a deal site may want to obtain an improved prediction of deal performance early in the day (and perhaps adjust accordingly). We present data for 7am for 2 reasons. First, Groupon generally sends email between 4am and 6am, and we want to see effects after the e-mail is sent. Second, as shown in Figure 6, on average Groupon deals have sold less than 7% of their eventual *first* day sales by 7am, so the time is suitable for an early prediction. Unsurprisingly, the predictive power of our model is greatly improved with this feature, as indicated by an increased $R^2$ of 0.81, and as shown in Figure 5. Detailed regression results are shown in Table 8a in the Appendix.

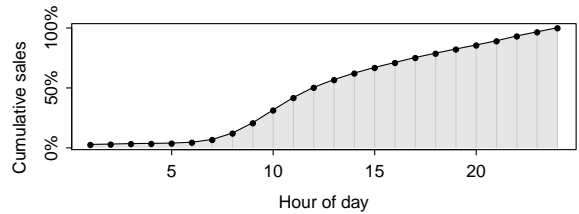

Figure 6: Average cumulative first day sales across all Groupon deals in our dataset.

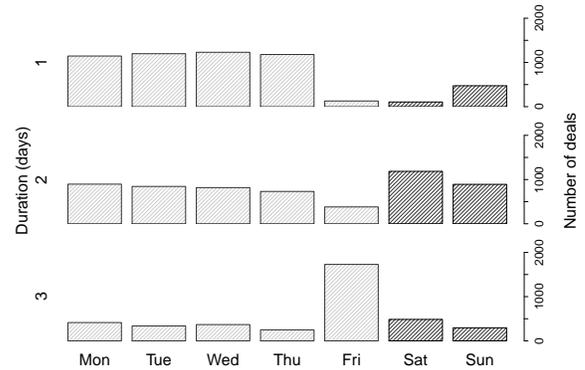

Figure 7: The number of deals starting on any given day of the week, broken down by their eventual duration.

**Yelp merchant reviews:** We also attempted to add the Yelp rating of merchants prior to their Groupon deal as a parameter to our model for the deals for which we had a rating available (2,522 of them). This failed to significantly improve our model. One explanation is that the average Yelp star-rating displayed for each business may not be especially significant to Groupon buyers; instead they might prefer scanning individual reviews and ratings on Yelp. Another is that there is only moderate variance in the merchants Groupon selects to give Yelp ratings for; we found 68% of the reviewed merchants had an average rating ranging between 3.5 and 4.5 stars, and 95% between 3 and 5 stars. Detailed regression results are shown in Table 8b in the Appendix. As a consequence of limiting our dataset to only on those merchants for whom we could acquire a Yelp rating we cannot obtain coefficient estimates for every deal category. For example, as we could not obtain any reviews for deals in the "Real Estate" category we cannot compute the corresponding coefficient.

## 3.1 Deal Scheduling

One further optimization available to daily deal sites is managing how deals are *scheduled*. Our data shows that Groupon is already managing deal schedules in non-trivial

**Following category**

|  | Art | Aut | Bea | Edu | Foo | Hea | Hom | Nig | Oth | Pet | Pro | Pub | Res | Sho | Tra |
|---|---|---|---|---|---|---|---|---|---|---|---|---|---|---|---|
| Art | 237 241 | 10/15 | 107/106 | 30/31 | 48/51 | 79/89 | 11/13 | 25/19 | 15/20 | 0/1 | 34/29 | 1/1 | 187/167 | 95/97 | 40/38 |
| Aut | 15/15 | 0 1 | 9/7 | 2/2 | 4/4 | 8/7 | 2/1 | 0/1 | 2/1 | 0/0 | 1/2 | 0/0 | 11/11 | 5/6 | 2/2 |
| Bea | 111/106 | 7/8 | 21 51 | 13/14 | 24/25 | 48/46 | 7/6 | 12/8 | 7/10 | 0/0 | 12/15 | 0/0 | 99/77 | 45/46 | 23/16 |
| Edu | 33/32 | 2/2 | 7/14 | 5 4 | 9/7 | 13/12 | 1/1 | 2/2 | 4/3 | 0/0 | 3/4 | 0/0 | 27/23 | 9/13 | 7/5 |
| Foo | 57/49 | 6/4 | 26/25 | 5/6 | 7 12 | 19/21 | 4/3 | 2/4 | 10/5 | 1/0 | 6/7 | 1/0 | 31/39 | 22/24 | 9/8 |
| Hea | 101/91 | 8/6 | 36/46 | 9/12 | 15/21 | 39 39 | 2/5 | 10/8 | 10/8 | 0/0 | 11/13 | 0/0 | 83/69 | 37/37 | 8/14 |
| Hom | 10/12 | 1/1 | 10/6 | 1/1 | 2/3 | 6/5 | 0 1 | 1/1 | 1/1 | 0/0 | 0/2 | 0/0 | 7/9 | 4/5 | 3/2 |
| Nig | 23/19 | 2/1 | 10/9 | 2/2 | 3/4 | 8/7 | 2/1 | 0 2 | 2/2 | 0/0 | 2/2 | 0/0 | 11/13 | 5/7 | 3/3 |
| Oth | 12/19 | 0/1 | 4/10 | 6/2 | 10/5 | 6/8 | 3/1 | 3/2 | 0 2 | 0/0 | 7/3 | 0/0 | 19/14 | 9/8 | 0/4 |
| Pet | 1/0 | 0/0 | 0/0 | 0/0 | 0/0 | 0/0 | 0/0 | 0/0 | 0/0 | 0/0 | 0/0 | 0/0 | 0/0 | 2/1 | 0/0 |
| Pro | 17/28 | 2/2 | 14/15 | 3/4 | 10/6 | 7/12 | 1/2 | 1/2 | 5/3 | 0/0 | 4 4 | 1/0 | 18/23 | 25/13 | 8/5 |
| Pub | 2/1 | 0/0 | 0/0 | 0/0 | 0/0 | 0/0 | 0/0 | 0/0 | 0/0 | 0/0 | 0/0 | 0 0 | 1/1 | 1/0 | 0/0 |
| Res | 177/167 | 16/11 | 114/77 | 26/23 | 35/39 | 83/70 | 7/9 | 8/13 | 10/14 | 1/0 | 24/22 | 1/1 | 65 128 | 83/71 | 21/25 |
| Sho | 95/98 | 6/6 | 50/46 | 14/13 | 28/22 | 37/38 | 4/6 | 4/7 | 10/9 | 1/1 | 12/13 | 0/1 | 72/70 | 37 44 | 14/14 |
| Tra | 29/38 | 1/2 | 20/16 | 6/5 | 12/8 | 12/14 | 4/1 | 3/3 | 3/3 | 0/0 | 2/5 | 0/0 | 36/25 | 10/14 | 4 6 |

(Rows labeled "Preceding")

Table 4: The observed/expected number of times one featured category follows another.

ways. For example, there are far fewer deals offered on weekends, but more multi-day deals run over weekends. In another example, in Figure 7, we plot the duration of deals by the day of the week that they started on. The subplots show the counts of one-day, two-day, and three-day deals by their starting day of the week. We observe that Groupon initiates most one-day deals Monday through Thursday and most three-day deals on Fridays and weekends.

We also find evidence that deals are scheduled according to their category; specifically, that Groupon avoids placing featured deals of the same category back-to-back over consecutive days in the same city. This is a natural strategy to maintain user interest, related to issues of ad fatigue (or advertising wear out) studied in other contexts [4, 6]. Table 4 summarizes an experiment we conducted to investigate this phenomenon. For each pair of categories we show the observed number of occurrences of that event followed by the expected number of occurrences if deals had been scheduled independently at random. Deviations from random are generally small, *except* for certain categories on the diagonal of this matrix, which indicates that Groupon avoids featuring the same category back-to-back.

We note that the problem of how deal sites should best schedule deals to optimize revenue or sales remains an interesting direction for future work.

## 4. THE IMPACT OF SOCIAL INFLUENCE

We now move from studying solely the deal data to considering the interplay between social networks and daily deal sites, specifically quantifying the impact of Facebook users while a deal is running, and of Yelp reviewers as they provide feedback about merchants.

We focus on two distinct issues. First, while many individuals may be enticed to purchase a Groupon based just on the daily e-mail, we hypothesize that many others are activated based entirely or in part by social factors. One observable quantity we leverage to investigate this hypothesis is the number of individuals who "like" a given deal on Facebook. Our measurement study quantifies the correlation between two sets of time-series: Groupons sold and Facebook likes. Our evidence indicates clear network effects from social influence, and is consistent with basic cascade models of activation from the theoretical literature.

Second, we consider another social construct: the influence of a Groupon sale on merchants' reputations as determined by online reviews. Specifically, we study Yelp reviews of restaurants before and after a Groupon offer. Not surprisingly, we find a surge in reviews from new customers subsequent to the offer. But we also find that reviewers mentioning "Groupons" and "coupons" provide strikingly lower rating scores than those that do not, and these reviews reduce Yelp scores over time in our dataset.

### 4.1 Do Deals Propagate via Facebook Likes?

Groupon deals can be shared with friends by text, e-mail, Facebook, Twitter, and other means. Facebook offers a readily observable measure, which we use as a proxy for more general social sharing here: Groupon has added a Facebook Like button on each deal's web page. Should a Facebook user decide that he likes a Groupon deal (no purchase necessary) and clicks the Like button, a like counter shown next to the button is incremented, and a short message to that effect is distributed to the user's Facebook friends by means of their news feeds. These messages may then propagate recursively through Facebook, potentially creating a sales-enhancing word-of-mouth effect.

Here we examine whether Facebook likes correlate with the final deal size, and then ask whether the data appears consistent with current models for the effects of social networks on buying decisions. We note recent work also suggests social spreading as a significant determinant of final deal size, giving a model where once the tipping point of a deal is reached, it grows due to a stochastic process that models social spreading [21]. Their model appears orthogonal to our analysis.

We emphasize that we must be careful not to conflate correlation and causation; it is not clear that likes inspire purchases, even if the number of likes and deal sizes are correlated. As a strawman, consider a model where each purchaser ignores prior likes, and pushes the Like button with probability $p$ after deciding to purchase. In this setting, likes and purchases would be closely (linearly) correlated, even though likes would not promote purchases. Indeed, this provides an interesting null hypothesis: are likes simply proportional to sales? Our data suggests that this is not the case: Figures 8a and 8b suggest a non-linear relationship, where the deal size is roughly proportional to the number of likes raised to a power much less than 1. Also, deals with approximately the same number of likes can vary significantly in terms of final deal size. (A more detailed breakdown appears in Figure 9.)

We can gain some further insight by adding likes to our regression model. Table 9 in the Appendix summarizes the results of regressing deal size on the various deal features and Facebook activity. Adding the logarithm of the number of likes as an additional variable on the right hand side of the model Equation 1 improves our $R^2$ statistic from 0.49 to 0.63 for Groupon and from 0.38 to 0.67 for LivingSocial.

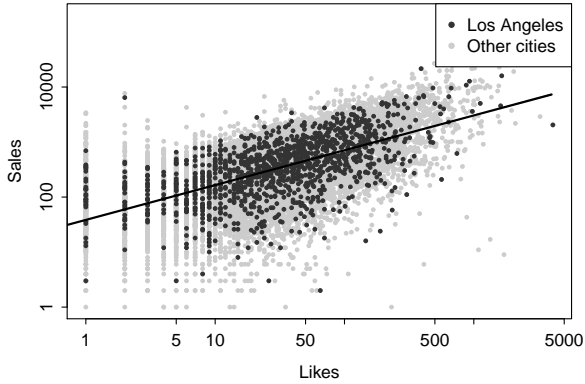
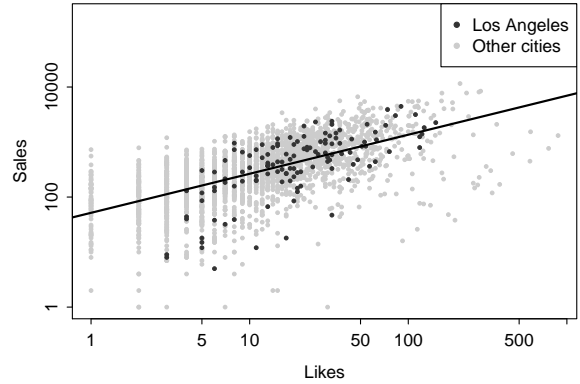

(a) Groupon, slope = 0.63  (b) LivingSocial, slope = 0.71

Figure 8: Deal size vs. number of Facebook likes (in logarithmic scale.) Black dots highlight deals in LA; grey dots are used for other cities. A trend line fitted using OLS regression is also shown.

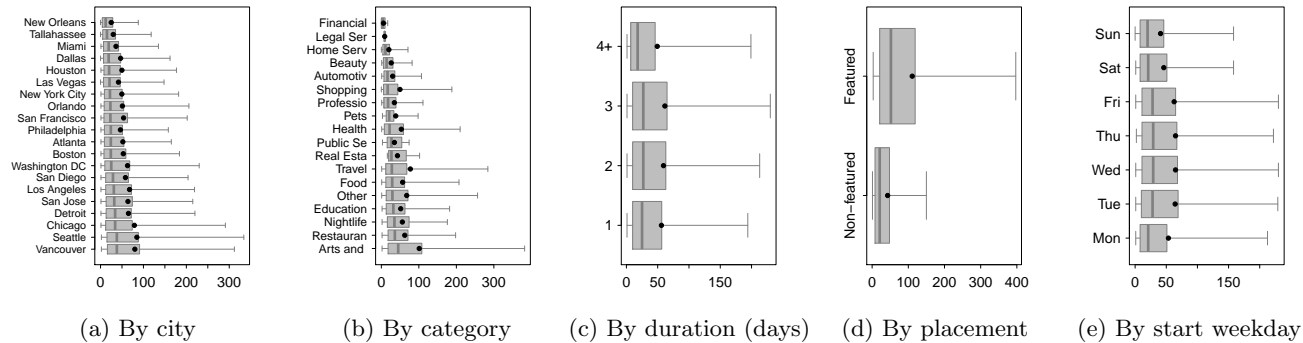

(a) By city  (b) By category  (c) By duration (days)  (d) By placement  (e) By start weekday

Figure 9: Number of Facebook likes are depicted on the $x$−axis. On the $y$−axis deals are broken down by various deal features. The outer whiskers mark the 5% and 95% quantiles of the deal size and revenue, the sides of each grey rectangle the 1st and 3rd quartiles, the solid bar the median, and the solid dot the mean.

While this improvement is not directly useful for, for example, predicting deal size *a priori*, it demonstrates a strong correlation between the logarithms of likes and deal size.

**Social Spreading:** We do not currently have (and have not sought) *direct* evidence of social spreading of deals via likes; this would require either detailed knowledge of the Facebook social network, or a detailed user study. Both are beyond the scope of this paper, but are interesting questions for future work. Here, however, we examine whether our data is consistent with theoretical models of social spreading from the literature (e.g., [13, 16]). In such models, there is generally a *seed set* of users that initially recommend a product; then additional users *activate* and buy the product based on these recommendations. Here, we treat Facebook likes as recommendations for a specific deal.

Two key features of cascade models are how the seed set is selected, and how inactive users are activated by neighbors. In our setting, we consider the seed set to be Groupon subscribers who are informed of the daily deal through Groupon's daily email and proceed to like the deal on Facebook. (Likely, those in the seed set also purchase the deal, but this is irrelevant to the cascade dynamics we consider here.) Our data provides insight into the size of the seed set, but not into the how the seed set should be selected. Previous work [5, 19, 20] has established the correlation between degree and activity in social networks. In [20] the authors demonstrate that the top 50% of Facebook users by degree are responsible for most social interactions. Kwak et al. [14] make a similar case for Twitter, where they demonstrate that users with more followers are more like to tweet. The potential implication, translated to our setting, is that Groupon customers who have more friends on Facebook are more likely to belong to the seed set. We test this by simulating cascade models with the seed set selected uniformly at random, and as the top-$k$ nodes by degree.

With regard to activation, we consider two standard variations: a user is activated with fixed probability $p$ by each of his active neighbors, and each active user $u$ activates user $v$ with probability $1/d_v$, where $d_v$ is the degree of user $v$. The first model activates high-degree nodes more frequently. These cases correspond to the Independent Cascade (IC) and Weighted Cascade (WC) models of Kempe et al. [13].

Finally, we model that each activated user purchases the deal with fixed probability $q$, which can be thought of as the conversion rate. Note that we separate the issues of activation and purchase; in our setting, activation corresponds to noticing the deal, rather than purchasing it.

We ran experiments based on these models on a network that has common characteristics with social network graphs, the arXiv High Energy Physics collaboration network from [15]. This network consists of 9,877 nodes and 51,971 edges. The results are shown in Figure 10. We plot the number of sales resulting by the cascade against the seed set size averaged over 100 trials per starting seed set size. We ran our experiments both by selecting a random seed set, and by selecting the top-$k$ nodes by degree. The high-degree heuristic resembles our empirical findings more closely. For the IC model we set $p = 2\%$, and for both models $q = 5\%$. We observe that in both cases, when using top-$k$ nodes as the seed set, we observe sublinear gains in the size of the cascade as the size of the seed set increases. These results match previous work such as [13], and give some insight into our empirical findings.

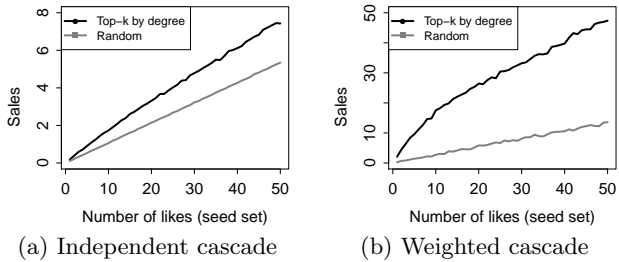

(a) Independent cascade  (b) Weighted cascade

Figure 10: Sales as predicted by two diffusion models.

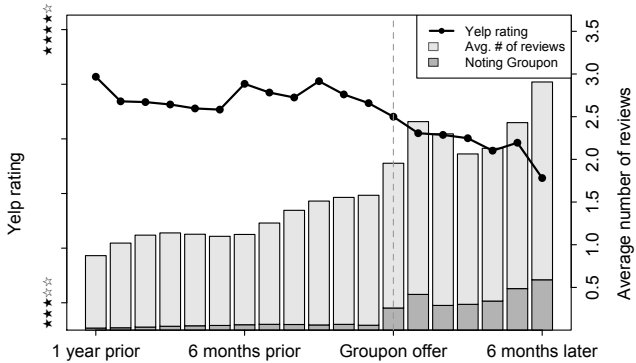

Figure 11: The average Yelp star-rating for merchants before and after their Groupon offers (line-chart), and the average number of reviews per merchant per month (bar-chart).

While these results suggest that social spreading is a feature of daily deals, current theoretical models are too simplistic to capture the array of features we have found to be important in determining deal size, with price being the most obviously relevant. It may also be useful to specifically consider models for daily deals in the context of cascading recommendations (e.g., [16]); this could also be helpful in explaining the large observed variance in deal sizes across deals with the same number of likes.

## 4.2 Yelp Feedback on Groupon Merchants

A key selling point of a daily deals site is the promise of beneficial long-term effects for merchants participating in a deal offering. Since discounted deals typically result in a net short-term loss to the merchant as customers redeem the coupons, a merchant is pitched on the expectation that some new customers, initially attracted by the deal, will become repeat customers, providing a long-term gain [9]. Participating merchants should determine that these gains outweigh the costs, which include providing discounts to their existing customer base.

|  | Avg. Rating | Reviews |
|---|---|---|
| Before period (12 months) | 3.71 | 39,042 |
| After period (6 months) | 3.59 | 17,006 |

Table 5: Yelp reviews around a merchant's Groupon offer.

While we do not have data to directly evaluate the long-term financial impact for Groupon merchants, we consider a novel, alternative approach to concretely quantify the impact of Groupon deals on a merchant's *reputation*. Specifically, we examine the extent to which a Groupon deal affects review scores at Yelp, a popular online review site. We view review scores as a useful proxy for both direct repeat business as well as for new business from word-of-mouth effects.

Groupon often displays a Yelp rating for the merchant offering a specific deal. For each Groupon deal associated with a Yelp rating, we collected all the individual reviews posted on Yelp up through August 2011. Yelp reviews are comprised of a star-rating ranging from one to five stars, the review text, and the date the review was written. We associated each review with an offset, measured in months, from the earliest Groupon deal in our dataset for the corresponding merchant, i.e., if a Groupon offer occurred in June 2011, a review for that merchant dated March 2011 would have an offset of -3. For each merchant, and for each integer offset (up through July 2011), we computed the merchant's average star-rating, thereby constructing a time-series of the star-rating value oriented around the first Groupon offer that we observed for a merchant. Figure 11 presents our findings. The line-chart displays the average star-rating across all Yelp-rated merchants in our dataset for each offset, with the $x$-axis depicting the offset from the time of the offer. The bar-chart depicts the average review volume over the same period, using the scale on the right-hand side of the y-axis, and measured in reviews per merchant per month. Lighter shading indicates total review volume, while darker shading indicates the volume of reviews that contain the keyword "Groupon". (Mentions of Groupon in reviews with negative offsets are mostly due to user reviews for merchants who ran an additional offer before we began collecting data. Excluding these merchants does not materially change our findings.)

In looking at Figure 11, we first consider the behavior *prior* to Groupon offers. While month to month scores vary, they appear essentially flat. The average number of reviews slowly increases (likely as Yelp itself is growing). There are few Groupon mentions. We think of these qualitative behaviors as our baseline.

Subsequent to the Groupon deal, we see marked changes in behavior. First, on average, the number of user-contributed reviews increases significantly after a Groupon offer, and based on the proportion of Groupon mentions, the Groupon deal appears to be the proximate cause. We quantify this change using the monthly number of Yelp reviews as a proxy. Our methodology is as follows: let $Y_i$ be the average number of reviews merchant $i$ received per month in the three months before the Groupon offer. Let $Z_i$ be the number of reviews the month of the Groupon offer. We compute the percentage change in number of reviews for each merchant the month of the Groupon offer as $(Z_i - Y_i)/Y_i$. (This requires $Y_i > 0$.) Similarly, we can compute the percentage change for the month after the Groupon offer. We find that the average percentage increase in reviews across all merchants who had received at least one review in the 3 months prior to the Groupon offer is 44%. Meanwhile, the average month-over-month growth in number of reviews for deals prior to their Groupon offers is about 5%. Similarly, the average percentage increase in reviews the month after the Groupon offer is 84%. Roughly 20% of merchants in our dataset (461 out of 2,332) had received zero reviews in the 3 months prior to the Groupon offer. Of these, 270 received at least one review within two months after the Groupon offer.

Our second conclusion is that Yelp star ratings decline after a Groupon deal. To quantify the magnitude of the decline, we employed the following methodology: as our baseline, we computed the average of all reviews with a negative offset (before the Groupon offer). Similarly, we computed the average of all reviews with a positive offset (after the Groupon offer). We deemed reviews with offset zero as ambiguous, due to data collection granularity, and they were not considered. The results of the before and after comparison are depicted in Table 5. The average drop in ratings is 0.12. This could affect any sorted order produced according to Yelp rankings significantly. Also, Yelp scores are reported and displayed according to discretized half-star increments. Thus, an average drop of 0.12 suggests a significant number of merchants may lose a half-star due to rounding. This could have a potentially important effect on a business; a recent study reports that for independent restaurants a one-star increase in Yelp rating leads to a 9% increase in revenue [17]. However, the transitory nature of Groupon-driven reviews, in addition to complexities of modeling hidden factors like weighted moving averages, cloud our ability to pinpoint the reputational ramifications precisely.

To provide further attribution for the decline, we conducted additional text analysis on the content of individual reviews. The results are summarized in Table 6, which categorizes the user-contributed reviews according to occurrence of the keywords "Groupon" and "coupon". Reviews mentioning either keyword are associated with star ratings that are 10% lower on average than reviews that do not, while the very small fraction of reviews mentioning both keywords are more than 20% lower on average.

Ultimately, the economic ramifications of reputational effects due to running a daily deal remain uncertain. The positive impact of quickly reaching a broad, new audience is precisely in line with the daily deals sales pitch, and is borne out by the surge in reviews that we witness in our dataset. However, the lower-on-average rating scores in those reviews mentioning Groupon provides a cautionary note: this could indicate that a more critical audience is being reached, or that the fit between the merchant and these new customers is more tenuous than with existing customers.

## 5. CONCLUSION AND FUTURE WORK

Our examination of daily deal sites, and particularly Groupon, has used data-driven analysis to investigate relationships between deal attributes and deal size beyond simple measures such as the offer price. Indeed, the scope of our investigation goes well outside of deal sites, to consider Groupon's relationship with the larger electronic commerce ecosystem, including Facebook and Yelp. We believe we expose signifi-cant complexity in understanding behavior in these systems. In particular, predicting deal sizes in settings where price, a multiplicity of other deal parameters, as well as the potential for social cascades to affect the outcome, provides a clear, albeit difficult, challenge. We also suggest the mining of publicly accessible Internet data, such as content on review sites, to benchmark the success of deal sites for merchants, in terms of the effect on their reputation. Expanding this approach to other data sources (such as Twitter or blogs) could yield further insights.

|  |  | Coupon mentioned | |
|---|---|---|---|
|  |  | Yes | No |
| **Groupon** | Yes | 2.98 (354) | 3.36 (4,315) |
| **mentioned** | No | 3.36 (1,166) | 3.71 (50,213) |

Table 6: Yelp reviews broken down by mentions of the keywords "Groupon" and "Coupon". The average rating as well as the number of reviews (in parentheses) are shown.

While our focus here has been on data analysis, we believe our work opens the door to several significant questions in both modeling and optimizing deal sites and similar electronic commerce systems.

## 6. REFERENCES

[1] K. S. Anand and R. Aron. Group buying on the web: A comparison of price-discovery mechanisms. *Manage. Sci.*, 49(11):1546–1562, November 2003.
[2] A. Arabshahi. Undressing Groupon: An Analysis of the Groupon Business Model. Available at http://www.ahmadalia.com, Dec. 2010.
[3] J. W. Byers, M. Mitzenmacher, M. Potamias, and G. Zervas. A Month in the Life of Groupon. Technical report arXiv:1105.0903, arXiv.org, May 4, 2011.
[4] B. J. Calder and B. Sternthal. Television commercial wearout: An information processing view. *Journal of Marketing Research*, 17(2):pp. 173–186, 1980.
[5] H. Chun, H. Kwak, Y.-H. Eom, Y.-Y. Ahn, S. Moon, and H. Jeong. Comparison of online social relations in volume vs interaction: a case study of cyworld. In *Proc. of the 8th ACM SIGCOMM Conference on Internet Measurement*, pages 57–70, 2008.
[6] C. S. Craig, B. Sternthal, and C. Leavitt. Advertising wearout: An experimental analysis. *Journal of Marketing Research*, 13(4):pp. 365–372, 1976.
[7] R. Crane and D. Sornette. Robust dynamic classes revealed by measuring the response function of a social system. *Proceedings of the National Academy of Sciences*, 105(41):15649–15653, 2008.
[8] U. M. Dholakia. How effective are groupon promotions for businesses? Available at http://www.ruf.rice.edu/~dholakia, Sept. 2010.
[9] B. Edelman, S. Jaffe, and S. D. Kominers. To Groupon or Not to Groupon: The Profitability of Deep Discounts. Available at http://www.scottkom.com/research.html, 2010.
[10] Groupon, Inc. S-1 Filing. Filed with the U.S. Securities and Exchange Commission, June 2, 2011.
[11] Groupon Blog: One Last Post on the Super Bowl. Available at http://www.groupon.com/blog/, Feb. 2011.

[12] R. J. Kauffman and B. Wang. New buyers' arrival under dynamic pricing market microstructure: The case of group-buying discounts on the internet. *J. Manage. Inf. Syst.*, 18(2):157–188, October 2001.

[13] D. Kempe, J. M. Kleinberg, and É. Tardos. Maximizing the spread of influence through a social network. In *KDD*, pages 137–146, 2003.

[14] H. Kwak, C. Lee, H. Park, and S. Moon. What is Twitter, a social network or a news media? In *Proceedings of the 19th International Conference on the World Wide Web*, WWW '10, pages 591–600, New York, NY, USA, 2010. ACM.

[15] J. Leskovec, J. Kleinberg, and C. Faloutsos. Graph evolution: Densification and shrinking diameters. *ACM Trans. Knowl. Discov. Data*, 1(1), March 2007.

[16] J. Leskovec, A. Singh, and J. Kleinberg. Patterns of influence in a recommendation network. In *Advances in Knowledge Discovery and Data Mining*, volume 3918 of *LNCS*, pages 380–389. Springer-Verlag, 2006.

[17] M. Luca. Reviews, Reputation, and Revenues: The Case of Yelp.com. Available at http://people.bu.edu/mluca/, 2011.

[18] Techcrunch: Valentine's Day Bait & Switch: Groupon Must Avoid Becoming Just Another Useless Coupon Site. Available at http://t.co/o5kfveu, Feb. 2011.

[19] B. Viswanath, A. Mislove, M. Cha, and K. P. Gummadi. On the evolution of user interaction in Facebook. In *Proceedings of the 2nd ACM Workshop on Online Social Networks*, pages 37–42, 2009.

[20] C. Wilson, B. Boe, A. Sala, K. P. Puttaswamy, and B. Y. Zhao. User interactions in social networks and their implications. In *Proceedings of the 4th ACM European Conference on Computer Systems*, EuroSys '09, pages 205–218, New York, NY, USA, 2009. ACM.

[21] M. Ye, C. Wang, C. Aperjis, B. A. Huberman, and T. Sandholm. Collective attention and the dynamics of group deals. Technical report available at http://arxiv.org/abs/1107.4588v1, 2011.

# APPENDIX

## A. DETAILED REGRESSION RESULTS

|  |  | Est. | Std. Err. | Signif. |
|---|---|---|---|---|
| **Incentives** | (Intercept) | 5.19 | 0.07 | *** |
|  | log Price | -0.48 | 0.01 | *** |
|  | log Threshold | 0.56 | 0.01 | *** |
|  | Multi-day? | 0.22 | 0.02 | *** |
|  | Featured? | 0.88 | 0.02 | *** |
|  | Limited? | 0.01 | 0.02 |  |
| **Start Day** | Mon | *Reference level* |  |  |
|  | Tue | 0.15 | 0.03 | *** |
|  | Wed | 0.22 | 0.03 | *** |
|  | Thu | 0.23 | 0.03 | *** |
|  | Fri | 0.24 | 0.03 | *** |
|  | Sat | 0.02 | 0.03 |  |
|  | Sun | -0.08 | 0.03 | ** |
| **Category** | Other | *Reference level* |  |  |
|  | Arts and Entertainment | -0.08 | 0.06 |  |
|  | Automotive | 0.12 | 0.08 |  |
|  | Beauty & Spas | 0.09 | 0.06 |  |
|  | Education | -0.21 | 0.07 | ** |
|  | Financial Services | -1.87 | 0.41 | *** |
|  | Food & Drink | 0.14 | 0.06 | * |
|  | Health & Fitness | -0.01 | 0.06 |  |
|  | Home Services | -0.28 | 0.08 | *** |
|  | Legal Services | -1.04 | 0.99 |  |
|  | Nightlife | -0.25 | 0.08 | ** |
|  | Pets | -0.33 | 0.13 | * |
|  | Professional Services | -0.17 | 0.07 | * |
|  | Public Services & Gov. | -0.61 | 0.38 |  |
|  | Real Estate | 0.14 | 0.50 |  |
|  | Restaurants | 0.22 | 0.06 | *** |
|  | Shopping | -0.02 | 0.06 |  |
|  | Travel | 0.14 | 0.07 | . |
| **City** | Atlanta | *Reference level* |  |  |
|  | Boston | 0.07 | 0.04 | . |
|  | Chicago | 0.32 | 0.04 | *** |
|  | Dallas | -0.02 | 0.05 |  |
|  | Detroit | -0.10 | 0.05 | * |
|  | Houston | -0.10 | 0.05 | * |
|  | Las Vegas | -0.03 | 0.05 |  |
|  | Los Angeles | 0.25 | 0.04 | *** |
|  | Miami | -0.27 | 0.05 | *** |
|  | New Orleans | -0.64 | 0.06 | *** |
|  | New York | -0.01 | 0.04 |  |
|  | Orlando | -0.21 | 0.05 | *** |
|  | Philadelphia | -0.37 | 0.05 | *** |
|  | San Diego | -0.08 | 0.05 |  |
|  | San Francisco | 0.24 | 0.05 | *** |
|  | San Jose | 0.06 | 0.06 |  |
|  | Seattle | 0.23 | 0.05 | *** |
|  | Tallahassee | -0.85 | 0.08 | *** |
|  | Vancouver | 0.24 | 0.06 | *** |
|  | Washington DC | 0.47 | 0.04 | *** |
|  | $F$-statistic | 345 | $p$-value | < 2e-16 |
|  | $R$-squared | 0.49 |  |  |
|  | Significance codes | 0% *** 0.1% ** 1% * 5% |  |  |

(a) Groupon

|  |  | Est. | Std. Err. | Signif. |
|---|---|---|---|---|
| **Incent.** | (Intercept) | 5.67 | 0.32 | *** |
|  | log Price | -0.73 | 0.03 | *** |
|  | Multi-day? | 0.34 | 0.05 | *** |
|  | Family edition? | -1.40 | 0.08 | *** |
| **Start Day** | Mon | *Reference level* |  |  |
|  | Tue | 0.11 | 0.07 |  |
|  | Wed | 0.22 | 0.07 | ** |
|  | Thu | 0.20 | 0.06 | ** |
|  | Fri | 0.28 | 0.08 | *** |
|  | Sat | -0.23 | 0.11 | * |
|  | Sun | -0.27 | 0.11 | * |
| **Category** | Other | *Reference level* |  |  |
|  | Beauty | 0.23 | 0.31 |  |
|  | Beer/Wine/Spirits | 0.11 | 1.05 |  |
|  | Body | 0.98 | 1.05 |  |
|  | Dental | -0.52 | 1.05 |  |
|  | Education | 0.32 | 0.49 |  |
|  | Entertainment | 0.11 | 0.31 |  |
|  | Escapes | 1.01 | 0.33 | ** |
|  | General | 0.78 | 1.05 |  |
|  | Getaway | 0.89 | 0.36 | * |
|  | Health | -0.05 | 0.31 |  |
|  | Home | 0.77 | 1.05 |  |
|  | Photography | 0.01 | 1.05 |  |
|  | Quick Bites | -1.31 | 1.05 |  |
|  | Restaurant | -0.05 | 0.31 |  |
|  | Retail | -0.23 | 0.31 |  |
|  | Services | 0.13 | 0.31 |  |
|  | Skincare | -0.53 | 1.05 |  |
|  | Specialty Food | -0.30 | 0.32 |  |
|  | Tour/Experience | 0.88 | 1.05 |  |
| **City** | Atlanta | *Reference level* |  |  |
|  | Boston | -0.51 | 0.12 | *** |
|  | Chicago | -0.36 | 0.11 | *** |
|  | Dallas | -0.12 | 0.13 |  |
|  | Detroit | -0.83 | 0.13 | *** |
|  | Houston | 0.17 | 0.13 |  |
|  | Las Vegas | -0.20 | 0.14 |  |
|  | Los Angeles | 0.46 | 0.13 | *** |
|  | Miami | -0.41 | 0.13 | ** |
|  | New Orleans | -0.31 | 0.15 | * |
|  | New York City | 0.45 | 0.11 | *** |
|  | Orlando | -0.38 | 0.13 | ** |
|  | Philadelphia | -0.25 | 0.12 | * |
|  | San Diego | -0.13 | 0.11 |  |
|  | San Francisco | 0.38 | 0.13 | ** |
|  | San Jose | -0.02 | 0.16 |  |
|  | Seattle | 0.52 | 0.13 | *** |
|  | Tallahassee | -1.31 | 0.15 | *** |
|  | Vancouver | 0.01 | 0.14 |  |
|  | Washington, D.C. | 0.96 | 0.13 | *** |
|  | $F$-statistic | 33.2 | $p$-value | < 2e-16 |
|  | $R$-squared | 0.38 |  |  |
|  | Significance codes | 0% *** 0.1% ** 1% * 5% |  |  |

(b) LivingSocial

Table 7: The regression model of deal sizes on their various features.

|  |  | Est. | Std. Err. | Signif. |
|---|---|---|---|---|
| Incentives | (Intercept) | 3.35 | 0.04 | *** |
|  | log Price | -0.07 | 0.01 | *** |
|  | log 7am Sales | 0.75 | 0.00 | *** |
|  | log Threshold | 0.16 | 0.01 | *** |
|  | Multi-day? | 0.33 | 0.01 | *** |
|  | Featured? | 0.30 | 0.01 | *** |
|  | Limited? | -0.11 | 0.01 | *** |
| Start Day | Mon | *Reference level* | | |
|  | Tue | -0.09 | 0.02 | *** |
|  | Wed | -0.04 | 0.02 | * |
|  | Thu | -0.05 | 0.02 | ** |
|  | Fri | -0.09 | 0.02 | *** |
|  | Sat | -0.07 | 0.02 | *** |
|  | Sun | 0.05 | 0.02 | * |
| Category | Other | *Reference level* | | |
|  | Arts and Entertainment | 0.16 | 0.04 | *** |
|  | Automotive | -0.11 | 0.05 | * |
|  | Beauty & Spas | -0.01 | 0.04 | |
|  | Education | 0.10 | 0.04 | * |
|  | Financial Services | -1.14 | 0.25 | *** |
|  | Food & Drink | -0.13 | 0.04 | ** |
|  | Health & Fitness | 0.10 | 0.04 | ** |
|  | Home Services | -0.04 | 0.05 | |
|  | Legal Services | -0.42 | 0.60 | |
|  | Nightlife | -0.07 | 0.05 | |
|  | Pets | -0.00 | 0.08 | |
|  | Professional Services | -0.05 | 0.04 | |
|  | Public Services & Gov. | 0.27 | 0.30 | |
|  | Real Estate | -0.11 | 0.30 | |
|  | Restaurants | -0.07 | 0.04 | * |
|  | Shopping | -0.06 | 0.04 | |
|  | Travel | 0.10 | 0.05 | * |
| City | Atlanta | *Reference level* | | |
|  | Boston | -0.08 | 0.03 | ** |
|  | Chicago | -0.21 | 0.03 | *** |
|  | Dallas | -0.12 | 0.03 | *** |
|  | Detroit | -0.13 | 0.03 | *** |
|  | Houston | -0.24 | 0.03 | *** |
|  | Las Vegas | -0.48 | 0.03 | *** |
|  | Los Angeles | -0.22 | 0.03 | *** |
|  | Miami | -0.11 | 0.03 | ** |
|  | New Orleans | -0.44 | 0.04 | *** |
|  | New York | -0.06 | 0.03 | * |
|  | Orlando | -0.17 | 0.03 | *** |
|  | Philadelphia | -0.21 | 0.03 | *** |
|  | San Diego | -0.35 | 0.03 | *** |
|  | San Francisco | -0.19 | 0.03 | *** |
|  | San Jose | -0.29 | 0.04 | *** |
|  | Seattle | -0.24 | 0.03 | *** |
|  | Tallahassee | -0.49 | 0.05 | *** |
|  | Vancouver | -0.14 | 0.04 | *** |
|  | Washington DC | 0.03 | 0.03 | |
| | $F$-statistic | 1330 | $p$-value | $< 2\text{e-}16$ |
| | $R$-squared | 0.81 | | |
| | Significance codes | 0% *** | 0.1% ** | 1% * 5% |

(a) The regression model of deal sizes on their features with the inclusion of early morning sales as an independent variable.

|  |  | Est. | Std. Err. | Signif. |
|---|---|---|---|---|
| Incentives | (Intercept) | 5.92 | 0.18 | *** |
|  | log Price | -0.47 | 0.03 | *** |
|  | Yelp Rating | -0.10 | 0.03 | *** |
|  | log Threshold | 0.57 | 0.02 | *** |
|  | Multi-day? | 0.19 | 0.05 | *** |
|  | Featured? | 0.70 | 0.05 | *** |
|  | Limited? | -0.04 | 0.04 | |
| Start Day | Mon | *Reference level* | | |
|  | Tue | 0.06 | 0.06 | |
|  | Wed | 0.08 | 0.06 | |
|  | Thu | 0.07 | 0.06 | |
|  | Fri | 0.04 | 0.06 | |
|  | Sat | -0.16 | 0.07 | * |
|  | Sun | -0.24 | 0.07 | ** |
| Category | Other | *Reference level* | | |
|  | Arts and Entertainment | -0.18 | 0.13 | |
|  | Automotive | 0.05 | 0.16 | |
|  | Beauty & Spas | -0.16 | 0.13 | |
|  | Education | -0.19 | 0.18 | |
|  | Food & Drink | -0.01 | 0.14 | |
|  | Health & Fitness | -0.07 | 0.13 | |
|  | Home Services | -1.07 | 0.25 | *** |
|  | Nightlife | -0.38 | 0.15 | * |
|  | Pets | -0.39 | 0.30 | |
|  | Professional Services | 0.07 | 0.17 | |
|  | Restaurants | 0.00 | 0.12 | |
|  | Shopping | -0.43 | 0.14 | ** |
|  | Travel | -0.01 | 0.20 | |
| City | Atlanta | *Reference level* | | |
|  | Boston | 0.18 | 0.11 | |
|  | Chicago | 0.38 | 0.10 | *** |
|  | Dallas | 0.33 | 0.13 | * |
|  | Detroit | -0.10 | 0.14 | |
|  | Houston | 0.09 | 0.13 | |
|  | Las Vegas | 0.24 | 0.13 | . |
|  | Los Angeles | 0.33 | 0.10 | ** |
|  | Miami | -0.37 | 0.14 | ** |
|  | New Orleans | -0.74 | 0.19 | *** |
|  | New York City | 0.08 | 0.10 | |
|  | Orlando | 0.16 | 0.16 | |
|  | Philadelphia | -0.44 | 0.12 | *** |
|  | San Diego | -0.09 | 0.12 | |
|  | San Francisco | 0.11 | 0.11 | |
|  | San Jose | 0.19 | 0.14 | |
|  | Seattle | 0.39 | 0.11 | *** |
|  | Tallahassee | -0.65 | 0.34 | . |
|  | Vancouver | 0.50 | 0.16 | ** |
|  | Washington DC | 0.74 | 0.12 | *** |
| | $F$-statistic | 64.8 | $p$-value | $< 2\text{e-}16$ |
| | $R$-squared | 0.53 | | |
| | Significance codes | 0% *** | 0.1% ** | 1% * 5% |

(b) The regression model of deal sizes on their features with the inclusion of Yelp rating as an independent variable.

Table 8: Two alternative regression models for Groupon deal sizes.

|  |  | Est. | Std. Err. | Signif. |  |  | Est. | Std. Err. | Signif. |
|---|---|---|---|---|---|---|---|---|---|
| Incentives | (Intercept) | 3.96 | 0.06 | *** | Incent. | (Intercept) | 3.52 | 0.24 | *** |
|  | log Price | -0.42 | 0.01 | *** |  | log Price | -0.57 | 0.02 | *** |
|  | log Likes | 0.42 | 0.01 | *** |  | log Likes | 0.82 | 0.02 | *** |
|  | log Threshold | 0.42 | 0.01 | *** |  | Multi-day? | 0.22 | 0.04 | *** |
|  | Multi-day? | 0.10 | 0.02 | *** |  | Family edition? | -0.97 | 0.06 | *** |
|  | Featured? | 0.58 | 0.02 | *** | Start Day | Mon | *Reference level* | | |
|  | Limited? | -0.01 | 0.01 |  |  | Tue | 0.06 | 0.05 | |
| Start Day | Mon | *Reference level* | | |  | Wed | 0.06 | 0.05 | |
|  | Tue | 0.07 | 0.02 | ** |  | Thu | 0.15 | 0.05 | ** |
|  | Wed | 0.12 | 0.02 | *** |  | Fri | 0.10 | 0.06 | |
|  | Thu | 0.11 | 0.02 | *** |  | Sat | 0.01 | 0.08 | |
|  | Fri | 0.21 | 0.02 | *** |  | Sun | -0.06 | 0.08 | |
|  | Sat | 0.10 | 0.03 | *** | Category | Other | *Reference level* | | |
|  | Sun | 0.01 | 0.03 |  |  | Beauty | 0.40 | 0.22 | . |
| Category | Other | *Reference level* | | |  | Beer/Wine/Spirits | 0.23 | 0.76 | |
|  | Arts and Entertainment | -0.22 | 0.05 | *** |  | Body | 0.10 | 0.76 | |
|  | Automotive | 0.32 | 0.06 | *** |  | Dental | 0.01 | 0.77 | |
|  | Beauty & Spas | 0.35 | 0.05 | *** |  | Education | -0.17 | 0.35 | |
|  | Education | -0.27 | 0.06 | *** |  | Entertainment | -0.19 | 0.22 | |
|  | Financial Services | -1.36 | 0.35 | *** |  | Escapes | -0.60 | 0.24 | * |
|  | Food & Drink | 0.17 | 0.05 | ** |  | General | 0.20 | 0.76 | |
|  | Health & Fitness | 0.02 | 0.05 |  |  | Getaway | -0.68 | 0.26 | ** |
|  | Home Services | 0.02 | 0.07 |  |  | Health | -0.01 | 0.22 | |
|  | Legal Services | -0.73 | 0.84 |  |  | Home | 0.31 | 0.76 | |
|  | Nightlife | -0.23 | 0.07 | *** |  | Photography | 0.20 | 0.76 | |
|  | Pets | -0.41 | 0.11 | *** |  | Quick Bites | -0.48 | 0.76 | |
|  | Professional Services | -0.06 | 0.06 |  |  | Restaurant | 0.08 | 0.22 | |
|  | Public Services & Gov. | -0.58 | 0.32 | . |  | Retail | -0.21 | 0.23 | |
|  | Real Estate | -0.18 | 0.42 |  |  | Services | 0.19 | 0.22 | |
|  | Restaurants | 0.25 | 0.05 | *** |  | Skincare | -0.29 | 0.76 | |
|  | Shopping | 0.10 | 0.05 | * |  | Specialty Food | -0.25 | 0.23 | |
|  | Travel | 0.03 | 0.06 |  |  | Tour/Experience | 0.52 | 0.76 | |
| City | Atlanta | *Reference level* | | | City | Atlanta | *Reference level* | | |
|  | Boston | 0.11 | 0.04 | ** |  | Boston | -0.05 | 0.09 | |
|  | Chicago | 0.24 | 0.03 | *** |  | Chicago | 0.16 | 0.08 | * |
|  | Dallas | 0.04 | 0.04 |  |  | Dallas | 0.09 | 0.10 | |
|  | Detroit | -0.17 | 0.04 | *** |  | Detroit | -0.52 | 0.10 | *** |
|  | Houston | -0.04 | 0.04 |  |  | Houston | 0.17 | 0.09 | . |
|  | Las Vegas | 0.07 | 0.04 | . |  | Las Vegas | -0.13 | 0.10 | |
|  | Los Angeles | 0.14 | 0.04 | *** |  | Los Angeles | 0.17 | 0.09 | . |
|  | Miami | -0.17 | 0.05 | *** |  | Miami | -0.15 | 0.09 | |
|  | New Orleans | -0.39 | 0.05 | *** |  | New Orleans | 0.10 | 0.11 | |
|  | New York | -0.00 | 0.03 |  |  | New York City | 0.44 | 0.08 | *** |
|  | Orlando | -0.18 | 0.04 | *** |  | Orlando | -0.30 | 0.09 | ** |
|  | Philadelphia | -0.27 | 0.04 | *** |  | Philadelphia | 0.02 | 0.08 | |
|  | San Diego | -0.04 | 0.04 |  |  | San Diego | 0.14 | 0.08 | . |
|  | San Francisco | 0.26 | 0.04 | *** |  | San Francisco | 0.57 | 0.10 | *** |
|  | San Jose | 0.08 | 0.05 |  |  | San Jose | 0.16 | 0.11 | |
|  | Seattle | 0.08 | 0.04 | * |  | Seattle | 0.44 | 0.10 | *** |
|  | Tallahassee | -0.59 | 0.07 | *** |  | Tallahassee | -0.55 | 0.11 | *** |
|  | Vancouver | 0.21 | 0.05 | *** |  | Vancouver | 0.45 | 0.10 | *** |
|  | Washington DC | 0.45 | 0.04 | *** |  | Washington, D.C. | 0.52 | 0.10 | *** |
| | $F$-statistic | 581 | $p$-value | < 2e-16 | | $F$-statistic | 110 | $p$-value | < 2e-16 |
| | $R$-squared | 0.63 | | | | $R$-squared | 0.67 | | |
| | Significance codes | 0% *** 0.1% ** 1% * 5% | | | | Significance codes | 0% *** 0.1% ** 1% * 5% | | |
| | (a) Groupon | | | | | (b) LivingSocial | | | |

Table 9: The regression model of deal sizes on their features with the inclusion of Facebook activity as an independent variable.